# *A smack of all neighbouring languages*: How multilingual is scholarly communication?

Carolina Pradier[1], Lucía Céspedes[1,2] and Vincent Larivière[1,2,3,4]

[1] École de bibliothéconomie et des sciences de l'information, Université de Montréal, Montréal, Québec, Canada.

[2] Consortium Érudit, Montréal, Québec, Canada

[3] Department of Science and Innovation-National Research Foundation Centre of Excellence in Scientometrics and Science, Technology and Innovation Policy, Stellenbosch University, Stellenbosch, Western Cape, South Africa.

[4] Observatoire des Sciences et des Technologies, Centre interuniversitaire de recherche sur la science et la technologie, Université du Québec à Montréal, Montréal, QC, Canada.

*Vincent Larivière.

**Email:** vincent.lariviere@umontreal.ca

**Abstract**

Language is a major source of systemic inequities in science, particularly among scholars whose first language is not English. Studies have examined scientists' linguistic practices in specific contexts; few, however, have provided a global analysis of multilingualism in science. Using two major bibliometric databases (OpenAlex and Dimensions), we provide a large-scale analysis of linguistic diversity in science, considering both the language of publications (N=87,577,942) and of cited references (N=1,480,570,087). For the 1990-2023 period, we find that only Indonesian, Portuguese and Spanish have expanded at a faster pace than English. Country-level analyses show that this trend is due to the growing strength of the Latin American and Indonesian academic circuits. Our results also confirm the own-language preference phenomenon—particularly for languages other than English—, the strong connection between multilingualism and bibliodiversity, and that social sciences and humanities are the least English-dominated fields. Our findings suggest that policies recognizing the value of both national-language and English-language publications have had a concrete impact on the distribution of languages in the global field of scholarly communication.

**Introduction**

Scholarly writing is often said to be an inherently polyphonic, dialogic exercise (1). Indeed, one of the skills every scholar must become increasingly proficient at is the incorporation of others' ideas and voices (2). Whether to reinforce an argument or to provide a counterpoint, the presence of others' words, craftily interwoven with one's own, are the defining feature of academic discourse. If science is built on cumulative knowledge, then the cited references present in publications are the material trace of this process.

Citations thus constitute one of the most relevant markers of symbolic capital within the scientific field (2–5). Citations are often considered as an indicator of impact, influence and quality, and are used in several research assessment systems, either at institutional, national or international level (6, 7). Since the importance of being cited cannot be overstated, it is not surprising that many studies have looked into citations as potential indicators of systemic inequities in science (8–10).

Language of publication is one source of such inequities, particularly among scholars whose first language is not English (11). In contrast to the markedly Anglophone monolingual circuits of mainstream science publishing and indexing (12–14), multilingualism in scholarly communications has been defined as the dissemination of research findings in all languages, arguing for equal access and equal valorization regardless of the language in which they are written (15). It is often linked to the concept of bibliodiversity (16, 17) and to efforts towards greater inclusion and acknowledgement of the science carried out and published in non-Anglophone contexts (18–20).

This article is aligned with such efforts and seeks to contribute to answering the overarching question of how much linguistic diversity there is in science. In going beyond the mere counting of the language of published scientific outputs, we aim to address an additional layer of multilingualism, namely, the place of languages as found in references of scientific publications. Previous research on linguistic diversity in science has mostly focused on specific regions and languages (16, 21–25). To the best of our knowledge, a single previous study by Garfield & Welljams-Dorof (26) conducted a global analysis of publication and citation languages across countries, but its main data source[1] has well-documented linguistic biases (12, 27). Using comprehensive data sources, we study the presence of languages across academic circuits of publications, in order to infer the extent to which different disciplines and regions tend to foster a mono or multilingual research landscape, as evidenced by the incorporation of multiple languages into their outputs (journals and publications) and inputs (cited literature).

**Results**

*To what extent is English the language of science?*

The proportion of English publications in Dimensions goes down from 93.86% in 1990 to 85.51% in 2023, mainly due to the growing percentage of Indonesian, Portuguese and Spanish documents (Figure 1A). While the number of papers published in English has consistently increased from 877,493 in 1990 to 4,989,675 in 2023, the gap separating it from all other languages of publication seems to be closing (see Supporting Information, Figure S2). Besides the growth in publications in other languages, this relative decrease of English can also be explained by an increase in indexing of non-English language journals, that is to say, the growing visibility of these scientific linguistic communities.

---

[1] The Institute for Scientific Information (ISI) database; today, Clarivate's Web of Science.

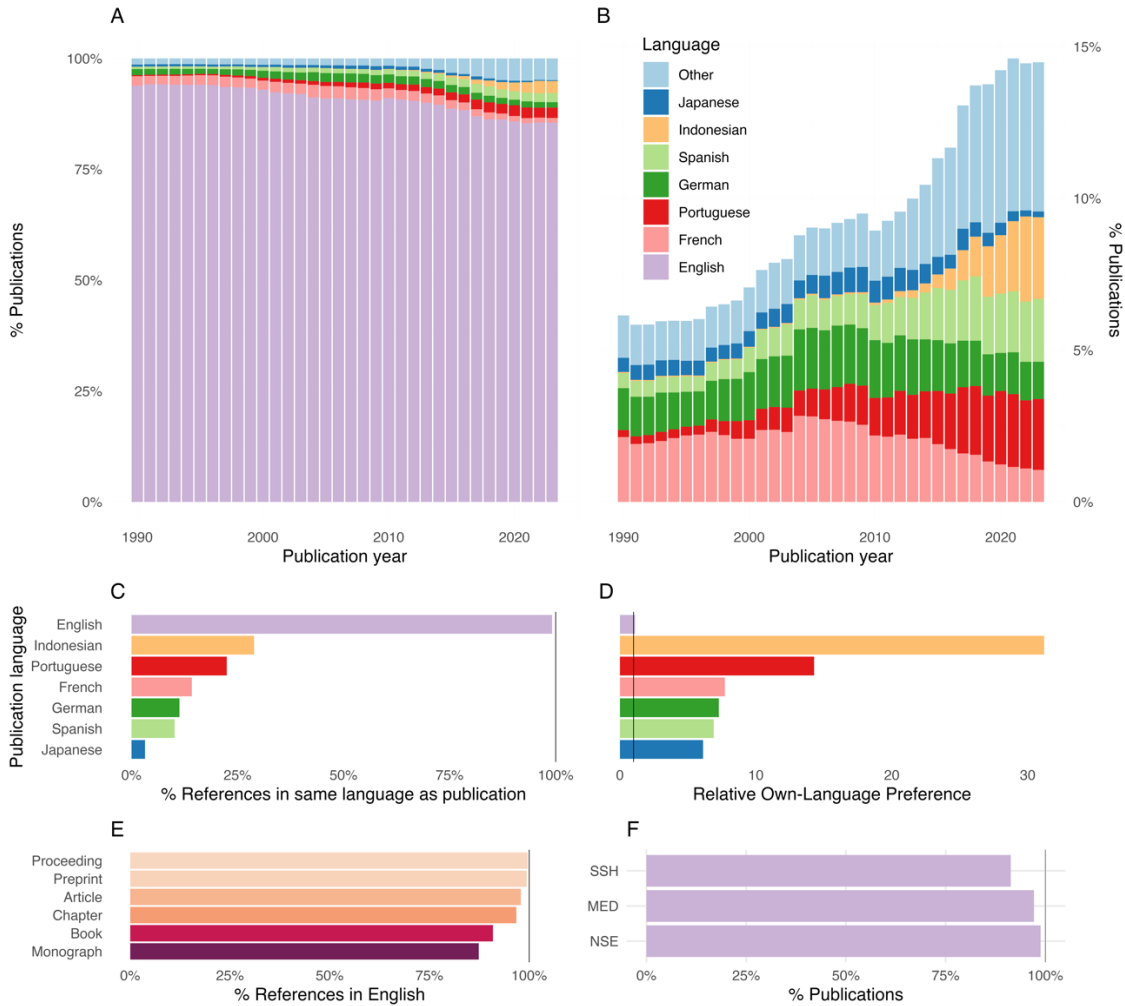

**Figure 1. Multilingualism at the article level (1990-2023).** A) Percentage of articles indexed in Dimensions by language. B) Percentage of articles indexed in Dimensions by language, excluding English. C) Percentage of cited references in the same language as the citing publication. D) Relative Own-Language Preference (ratio between the observed same language citation rate and the expected same language citation rate). E) Percentage of English-language references by document type. F) Percentage of publications in English, by discipline (SSH: Social Sciences and Humanities; NSE: Natural Sciences and Engineering; MED: Biomedical and Health Sciences).

The breakthrough of Indonesian is worth highlighting, from almost complete invisibility, to 2.69% of publications in 2023 (Figure 1B). In 2014, Indonesian universities were mandated by government decree to make all scientific publications openly available (28), a landmark that is reflected in our data. On the other hand, the two main Latin American languages show a gradual but sustained increase in publications, reaching similar proportions by the end of the period considered and surpassing, by far, central European languages. French arguably experienced the most significant fall, going from being the second largest language of scientific publication in 1990 to a mere 1.05% by 2023. German, after experiencing some growth in the early 2000s, has also lost ground in the last decade.

Although these results depict an increasingly multilingual portrait of science—or, at least, an improvement in bibliographic databases' capacity to provide a comprehensive portrait of science— English remains the overwhelmingly dominant language: 98.89% of all cited references in our corpus point to documents in English. However, our results confirm how closely linked multilingualism and bibliodiversity are, as the percentage of English-language references varies across document types (Figure 1E). While citations to conference proceedings, preprints, articles and even book chapters in English are well above 90% of all references to these document types, references to books and monographs refer to documents in English in 90.91% and 87.35% of cases respectively.

Our findings are also coherent with the phenomenon known as own-language preference (Figures 1C and 1D), that is, "the degree to which researchers in a certain field draw upon the literature published in their own language" (29)[2]. However, these results must take into account the language composition of citable documents. The Relative Own-Language Preference (ROLP) indicator represents the ratio between the *observed* same language citation rate and the *expected* same language citation rate—the language's share of publications (29). For instance, if 50% of language *L*'s references are to publications written in language *L*, but only 10% of the publications are written in *L*, we would obtain a ratio of 5.

English-language articles mainly cite literature in the same language (Figure 1C). However, this phenomenon is not due to a disproportionate preference to cite other English-language literature, but rather due to the overwhelming majority of citable documents in that language (Figure 1D). As for Indonesian, more than a quarter of the references found in articles written in this language are to other Indonesian publications, and this case does reflect an aggressive preference for articles written in Indonesian. In Latin America, both Portuguese and Spanish constitute their own circuits of regional scholarly publishing and knowledge circulation as well (13, 16, 30, 31), but own-language preference seems to be less pronounced for hispanophone researchers than for their lusophone counterparts.

The role of languages also varies across scientific disciplines due to differences in research cultures, as well as the nature of certain objects of study across academia. Figure 1F shows the proportion of publications in English for each of the three broad categories defined in the Methods section. While in NSE almost all documents are found to be written in English, this proportion is somewhat lower in MED. It is in SSH where we find the comparably lowest share of English-language publications; however, with a 91.33%, the supremacy of English in SSH publications can hardly be questioned.

*The role of journals as venues for non-English conversations*

Given their role as venues of circulation of knowledge, journals are key to structuring academic fields (32). According to our results, multilingual journals and journals published in languages other than English currently represent little more than 20% in MED and NSE, while this proportion is closer to 40% in SSH (Figure 2A). French and German still retain a minimal presence among MED and NSE journals, but it is in SSH where their decline has been more pronounced, particularly for

---

[2] A consideration is in order: Bookstrein and Yitzhaki regard own-language preference as a phenomenon of language speakers. In our interpretation, the coincidence between language of citing and cited documents may not involve the authors' *own* language, but rather, that in which they choose to write and seek to publish. At this point, it would thus be more precise to speak of *same*-language preference. However, we will refer to national and regional linguistic tendencies for scholarly publishing in section 3 of this article.

French. Japanese journals maintained a modest, if diminishing, proportion of journals in all three fields for most of the considered period, but sharply declined in the last two years.

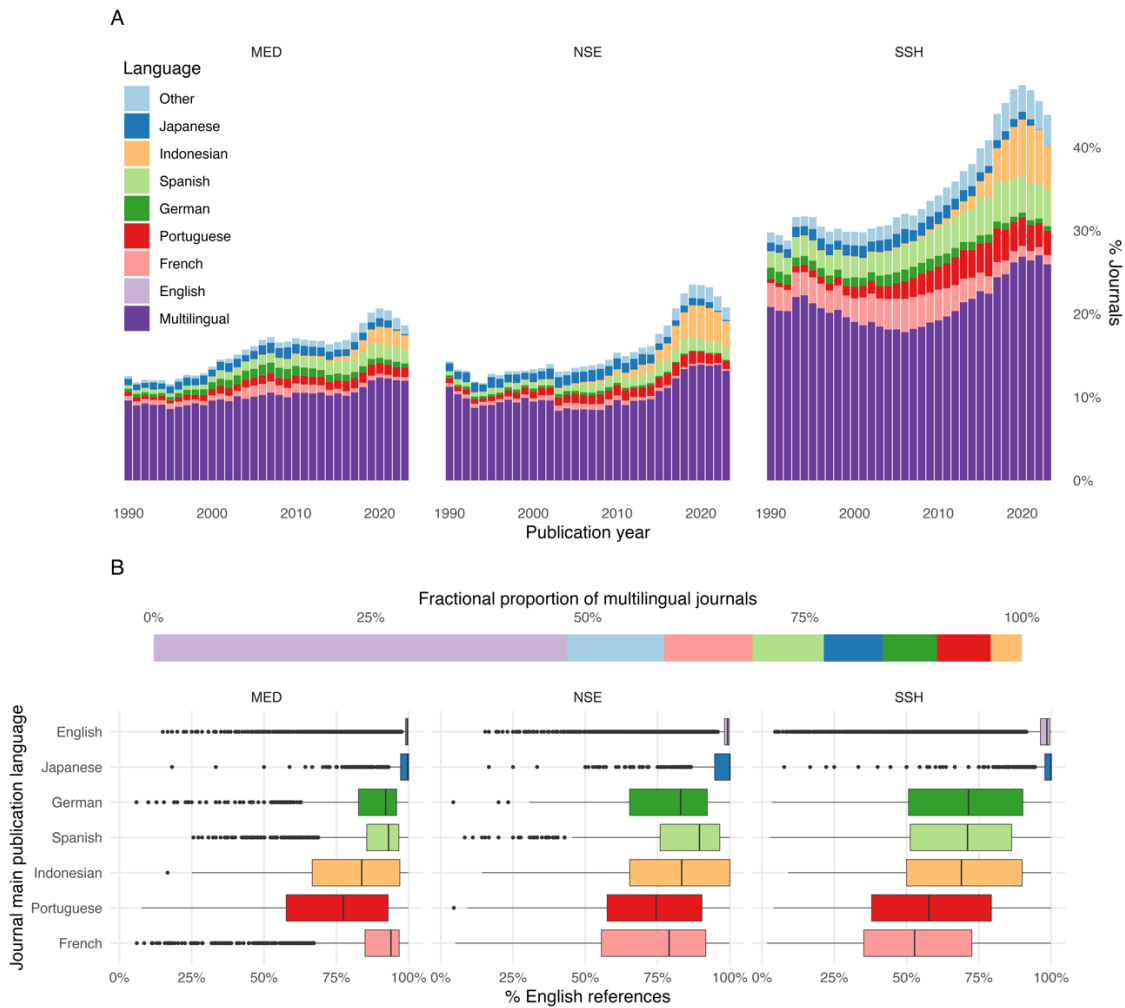

**Figure 2. Multilingualism at the journal level (1990-2023).** A) Percentage of journals indexed in Dimensions by language and discipline, excluding English. B) Language composition of multilingual journals. C) Distribution of English references by discipline and journal publication language.

Both Spanish and Portuguese journals have increased their shares in all three disciplinary fields. This growth can be attributed to the development of region-wide indexes and infrastructures for scholarly publishing and scientific information—such as Latindex (1995), Scielo (1998), Redalyc (2003) or LaReferencia (2012)—which have provided Latin American journals with the necessary support and tools to expand their activities. These platforms represent a fundamental system of conservation and visibility of Latin American research across disciplines (33). The trajectory of Indonesian journals is somewhat different, since its explosive growth is not built on pre existing academic circuits, but is driven by the massive adoption of the Open Journal System (OJS) as an infrastructure for digital publication over the last two decades, making the country the current world-leader in open access journals editing (34).

While the proportion of multilingual journals has remained fairly stable for most of the period in MED and NSE, and even began to experience some growth in the last five years (more markedly in the case of NSE), these journals went down in SSH from 22.23% to 17.8% between 1994 and 2006. The tendency has been reversed, reaching a higher percentage in 2023 than that of 30 years ago. This speaks of a shifting trend in the policies of SSH journals, increasingly accepting submissions and publishing articles in at least two languages.

In order to verify whether this multilingual turn is real or just a formal declaration of principles by journals, we examined the language distribution of the papers in multilingual journals (Figure 2B). As we can see, almost half of the documents published in journals that formally accept submissions in different languages are actually in English. French comes in second, with 10.17% of articles published in multilingual journals. The scarce 3.6% of articles in Indonesian published in multilingual journals may indicate a preference for publication in Indonesian-only journals when the articles are written in this language.

Unsurprisingly, almost all references found in English journals are to English documents (Figure 2C). This applies to Japanese journals as well, revealing—at least in terms of the available data—a linguistic community that has been completely absorbed by the English mainstream. For all other languages, we observe a greater dispersion in the language composition of citations, indicating varying levels of integration into the mainstream English circuit within each linguistic community.

We find the lowest rates of citations to English-language literature in articles in SSH journals, especially for French, where median citations to English documents barely surpass 50% of the total. There certainly are many prominent French SSH journals, but their sphere of influence is limited to their national and linguistic fields, while they remain invisible to the USA (35). This is indicative of a still closely-knit linguistic and academic community which reads and cites the knowledge that circulates in French, backed up by the symbolic capital held by some iconic French publications, editorials and institutions, by digital infrastructures such as libraries and repositories, and by linguistic policies seeking to maintain (or restore) the relevance of the French language in the scholarly communications landscape.

Conversely, articles published in NSE, and especially in MED journals, are more likely to reference English-language literature. In this regard, Portuguese and Indonesian journals stand out, as they have built strong linguistic and academic circuits of publication that operate in all fields—not only in SSH, where multilingualism is widespread. In the English-dominated fields of NSE and MED, this is no small feat.

These results highlight the tension between the two main functions of scientific journals: being a means of communication and an instrument of consecration (36, 37). Even though evaluation systems encourage scientists to publish their work in high-impact—which is often synonym to English-language—journals, non-mainstream journals in languages other than English provide essential venues for developing locally relevant scientific conversations and fostering the dissemination of scientific knowledge to the general public (9, 25, 38, 39).

*Where is multilingualism coming from? National and regional dissemination circuits*

The distribution of English-language publications across countries (Figure 3) exhibits roughly the same pattern as English-language references (Figure 4), but the bar is set much, much higher for references. Worldwide, even in the countries with the lowest proportions of articles in English, such as Indonesia, Brazil, Ecuador, or Angola, references to English publications represent no less than 92%.

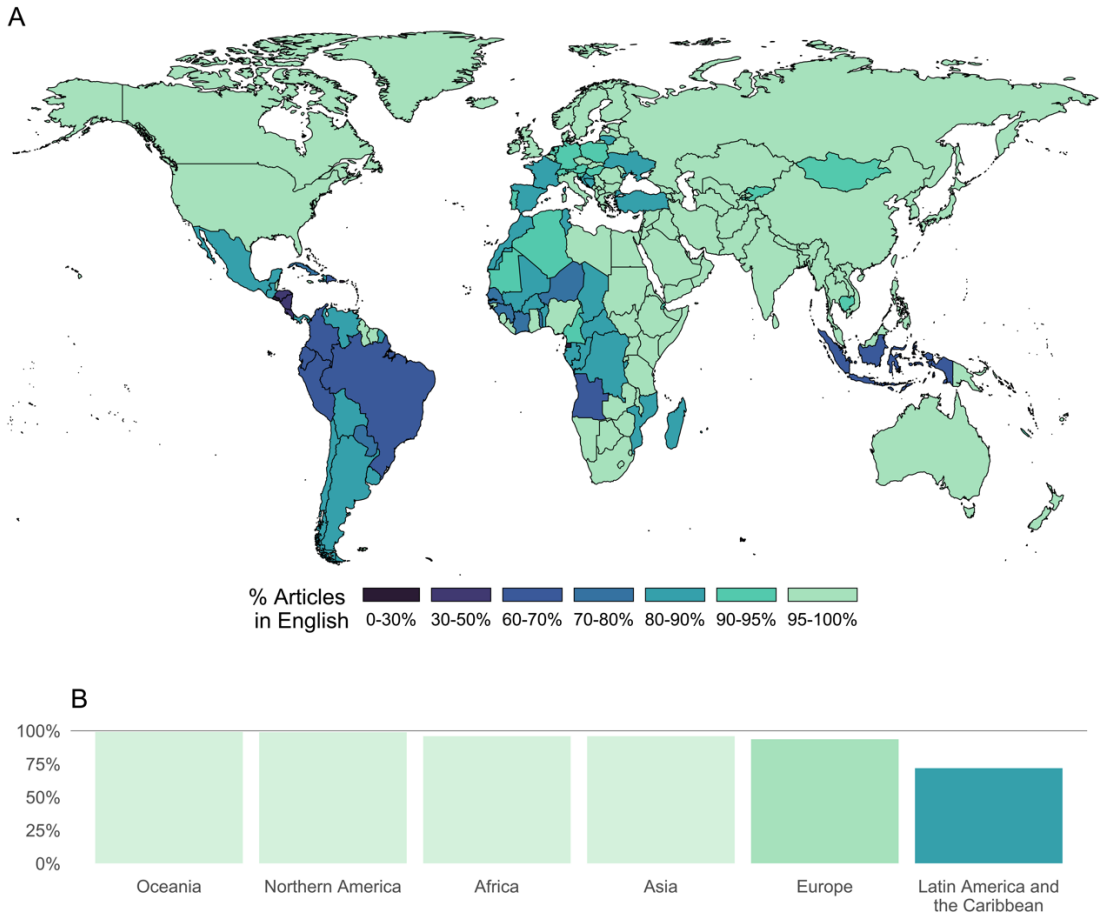

**Figure 3. Multilingualism in publications across the world (1990-2023).** Proportion of articles and conference proceedings in English by A) country (countries with at least 30 publications) and B) region (fractional counting).

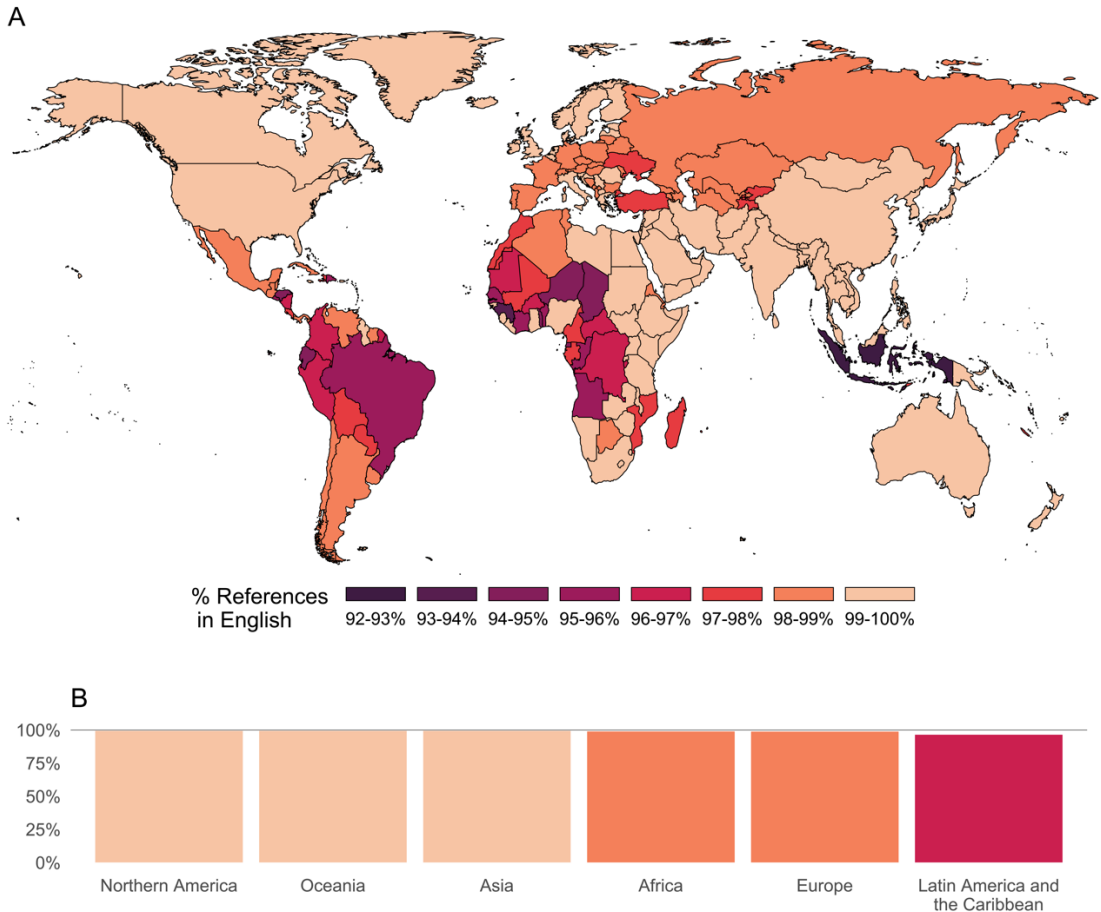

**Figure 4. Multilingualism in cited references across the world (1990-2023).** Proportion of references in English by A) country (countries with at least 30 publications) and B) region (fractional counting).

This world distribution of English publications and references corresponds almost exactly with Kachru's (40) model of concentric circles of world Englishes:

> These circles are defined with reference to the historical, sociolinguistic and literary contexts. (...) The Inner Circle represents the traditional bases of English, dominated by the 'mother tongue' varieties of the language. In the Outer Circle, English has been institutionalised as an additional language (...) The Expanding Circle includes the rest of the world where English is used as the primary foreign language, and the uses of English are unpredictably increasing. (p. 3)

As we can see, Inner-Circle Anglophone countries such as the USA, the UK, Ireland, Australia, and New Zealand practically do not publish in any other language than English. In the case of Canada, it can be inferred that the majority of English-speaking scholars' publication patterns drive the average of this country closer to an almost complete hegemony of English rather than towards a balanced bilingualism. As for references, these countries seem to be impervious to literature in languages other than English.

Outer-Circle African and Asian countries where English is an official or co-official language or has any other kind of legal status in public administration or education, such as Kenya, Nigeria, South Africa, or India, also lean towards English publications and references almost exclusively. On the contrary, in countries with a history of invasion and colonization by the French, Portuguese and Belgian empires—for example, Niger, Mozambique, or the Democratic Republic of Congo—, or where Arabic plays an important role in scholarly communications—for example, Morocco, Algeria, or Tunisia) (41) —English does not appear as the sole language of scholarly communication. These patterns reveal the deeply entrenched colonial legacy in the linguistic heritage and academic cultures of former European colonies, and highlight the political and epistemic efforts still ahead to decolonize the structures of production, circulation and appropriation of scientific knowledge (42–44).

Meanwhile, in the Expanding Circle, central non-Anglophone countries with strong academic traditions exhibit slightly lesser proportions of publications and references in English (mainly France and Spain, and, to a lesser extent, Germany). The case of Italy is particularly striking, as the language of Dante seems to have been completely overridden by English in both publications and references. Interestingly, Portugal shows a higher anglicization than these countries and is in the antipodes of Brazil. This is a different situation than that of Spain, which leans closer to the behavior of some Spanish-speaking Latin American countries, both in terms of language of publications and references. Countries in Eastern Europe and Asia—except for Indonesia, and to some extent, Turkey, Ukraine, and Russia—have a near complete preference to cite English-language documents. However, we can hardly draw robust conclusions from our results for Russian, Korean and Chinese-speaking countries, as OpenAlex performs particularly poorly in identifying publications in these languages (45).

Finally, Latin American countries consistently exhibit low rates of publications and references in English. However, intra-regional nuances can be observed: Brazil, Colombia, Ecuador and Peru publish in English markedly less than México, Venezuela, Chile, Argentina or Uruguay. Guyana and Suriname, where the official languages are English and Dutch respectively, are particular cases due to their political and cultural ties with the Anglophone Caribbean and the Netherlands. El Salvador, Nicaragua and Honduras, on the contrary, are the countries with the lowest proportion of English publications worldwide.

Focusing on the regional level aggregations (Figures 3B and 4B), Latin America stands out as the only region that has successfully developed a research dissemination ecosystem that is relatively detached from the English-speaking mainstream. While the Indonesian case presents a promising example of successful national-level policies aimed at countering English-language hegemony, its linguistic isolation within the region poses challenges for the future growth of an Indonesian language community. Africa may seem paradoxical at first glance, since the continent has some of the countries with the highest and lowest percentages of English-language references. However, this division clearly corresponds to the linguistic distribution of the European colonization of Africa, where clusters of countries with low percentages of English-language publications and references are basically still under the influence of colonial language communities that are gradually losing their power (46). Conversely, the characteristics of the colonization process in Latin America resulted in a landscape where almost all countries share a single official language, and the second most relevant language has common linguistic origins, resulting in low barriers to communication.

**Discussion**

Despite the dominance of English, the production of scientific knowledge is a multilingual endeavor. In the last two decades, we have witnessed—partly thanks to a better indexing—a change in the center of gravity from the European and North American "triumvirate" of English, French and German that inaugurated the XX century, and that gradually compressed into an almost complete Anglo hegemony (47), to an increasingly South American- and South Asian-leaning landscape of Indonesian, Portuguese and Spanish publication circuits. Our study brings attention to the fact that, if English acts both as a bridge and a fence in international exchanges of scientific knowledge (48), the dialogues happening under the bridge and around the fence should not be overlooked.

Important limitations to those results have to be acknowledged, however. They represent lower-bound estimates of the linguistic distribution of publications and citations. This is due to a number of factors. First, OpenAlex often misclassifies non-English texts as if they were in English (45), which leads to an overestimation of English in detriment to other languages. This means that, in reality, the research dissemination ecosystem is more multilingual than our results suggest. Second, citation links data are more limited for non-English publications. Third, bibliographic databases (even those with a broad coverage such as Dimensions or OpenAlex) do not provide a complete coverage of non-mainstream circuits, where many publications often do not even have a DOI. This leaves out of the picture a very relevant part of scientific outputs from regions or institutions that cannot afford to pay for DOIs in their publications (49). For example, as surveyed by Authier (50), 54% of Latin American journals indexed in the Directory of Open Access Journals lack any kind of persistent identifiers. As large-scale bibliometric databases such as OpenAlex and Dimensions rely on Crossref as its backbone, journals that use other DOI registration agencies[3] are less likely to be covered. As most of those agencies are in Asia, journals from that region may suffer from indexing issues, which results in inaccurate representations of these countries' languages in our maps. The Web of Science does have an Arabic, Chinese, and Russian Citation Database as part of their collections (in addition to integrating Scielo and the Korean Journal Database). But WoS has a more restrictive coverage than Dimensions and OpenAlex, therefore, these regional indexes might still be leaving out a sizable proportion of journals. In turn, those covered by wider-ranging databases may be lacking well-curated metadata for citations and languages.

English monolingualism can be understood in terms of what Law and Lin (51) call "market-oriented parochialism", this is, the demands from commercial publishers—particularly those based in North America and Europe—that publications be made more "international", where "international" means attuned to dominant research agendas. As Curry and Lillis (52) have observed, in recent decades the signifier "international" has slid towards the signifier "English", and together, both terms constitute a naturalized—but by no means natural—indicator of scientific quality. The linguistic communities we have identified in this study—mainly Indonesian or Brazilian scholars citing literature written in their own languages—emerge as spaces that question the assumption that publishing in one's own country is an endogamic and therefore discouraged practice (30), valuing, instead of marginalizing, "the ways of knowing and being entwined in these languages and the people and communities who embody, care for, and practice this knowledge" (19).

The fact that English-language publications and references still account for the vast majority does not imply a passive adoption or a blind incorporation of foreign theoretical frameworks, nor an uncritical alignment with Inner-Circle schools of thought. For non-Anglophone authors, publications or translations into English may also be a strategy to reach visibility in other national fields (53), thus potentially enabling South-South collaborations and dialogues facilitated by English as a vehicular or contact language[4]. Nevertheless, although a higher reliance on studies published in English does not automatically entail that such cited research has been entirely conducted in the Inner-Circle and thus responds to central research agendas, it does raise the question of the geographical and institutional concentration of resources needed to perform and publish the most influential research in each discipline worldwide, as well as issues of symbolic capital and prestige that favor the circulation, reception and adoption of certain schools of thought, theories and methodologies (55). In MED and NSE, the majority of English-language publications and references is a warning sign of an expanding epistemological monoculture (56). On the other hand, the so-called parochialism or provincialism of SSH may be saving epistemic and linguistic diversity in that field.

---

[3] https://www.doi.org/the-community/existing-registration-agencies/

As Bourdieu (55) has stated, no intellectual, cultural or scientific exchanges are spontaneously international; rather, they are marked by social operations of selection, demarcation and reception. So, at this point, we may correct the terms we ourselves used elsewhere in this article: the behavior of languages is none other than the behavior of their speakers. Thus, if Indonesian or Portuguese occupy a more prominent position today than they did 30 years ago, it is owing to the massive adoption of Open Journal Systems (an open-source, free-to-use journal publishing platform) or to the enactment of policies that value publications in the national language as well as those in English. Both are examples of science policy with a linguistic angle that have had a very concrete impact on the distribution of languages in the global field of scholarly communication, effectively contributing to multilingualism and bibliodiversity and, in the process, allowing other voices to be heard over the sometimes deafening and seemingly impregnable Anglophone wall of sound.

**Materials and Methods**

Data for this article were retrieved from the Dimensions and OpenAlex databases (57, 58). We examine all articles and conference proceedings indexed in Dimensions and published between 1990 and 2023. Language information—confirmed using manual validation of a sample (45)—was retrieved from OpenAlex, and matching between both bibliometric databases was based on Digital Object Identifier (DOI) matching. Throughout our results section, all languages below the threshold of 0.5% publications each are grouped into the category labeled as "Other". Russian was also included in this group despite exceeding the 0.5% threshold due to potential indexing issues at the sources of data, which rendered the results unreliable.

Our data consist of 87,577,942 distinct articles and conference proceedings with language information, 54,822,930 of which also included information on their references, resulting in a total of 1,480,570,087 citation links. Our results should be interpreted with the following caveat: publications with information on their references are not randomly distributed across languages, they are much more frequently found in English documents than in other languages (see Supporting Information, Figure S1). Therefore, all of our estimations of the degree of multilingualism of references should be considered as lower bounds.

Dimension's disciplinary classification is based on the Fields of Research (FoR) classification of the Australian and New Zealand Standard Research Classification (ANZSRC) system. This classification system includes 22 research disciplines, which were regrouped into three broad research areas (Table 1).

---

[4] We consciously avoid the denomination of "English as a *lingua franca*" due to the asymmetry of power involved: while English is arguably the most chosen language for communication between speakers of different linguistic backgrounds, it also is "the national language of the main hegemonic centers in the world-system, and the native language of some of the most influential agents in the global scientific field" (54).

**Table 1.** Reclassification of Australian and New Zealand Standard Research Classification into broad research areas.

| Broad research area | ANZSRC Fields of Research |
|---|---|
| Biomedical and Health Sciences (**MED**) | Biological Sciences, Biomedical and Clinical Sciences, Health Sciences, and Psychology |
| Natural Sciences and Engineering (**NSE**) | Agricultural, Veterinary and Food Sciences, Earth Sciences, Environmental Sciences, Information and Computing Sciences, Mathematical Sciences, Built Environment and Design, Chemical Sciences, Engineering, and Physical Sciences |
| Social Sciences and Humanities (**SSH**) | Commerce, Management, Tourism and Services, Creative Arts and Writing, Economics, Education, History, Heritage and Archaeology, Human Society, Language, Communication and Culture, Law and Legal Studies, Philosophy and Religious Studies |

Articles in our corpus were published in 107,259 distinct journals. Journal language was defined in terms of the most frequent language of publication. In cases where the most frequent language accounted for less than 90% of publications, the journal was considered multilingual. Finally, country-level authorship is computed using fractional counting. Each publication is divided by its number of authors, and these fractions are then assigned to each country according to each author's first institutional affiliation (in cases of multiple affiliations). These fractions are later aggregated to determine the proportion of the articles authored by each country (the sum of all fractions equals the number of publications in our dataset). While our country-level analysis cannot provide a detailed report for every single country, readers can explore an interactive version of our results at https://vlab.ebsi.umontreal.ca/languages_app/.

**Acknowledgments**

The authors acknowledge funding from the Social Science and Humanities Research Council of Canada Pan-Canadian Knowledge Access Initiative Grant (Grant Number 1007-2023-0001), and the Fonds de recherche du Québec—Société et Culture through the Programme d'appui aux Chaires UNESCO (Grant Number 338828).

**Author Contributions:** C.P., L.C., and V.L. designed research; C.P., L.C., and V.L. performed research; C.P. analyzed data; and C.P., L.C., and V.L. wrote the paper.

**References**

1. M. M. Bakhtin, *The Dialogic Imagination: Four Essays* (University of Texas Press, 2010).
2. K. Hyland, F. (Kevin) Jiang, Points of Reference: Changing Patterns of Academic Citation. *Appl. Linguist.* **40**, 64–85 (2019).

3. N. Desrochers, *et al.*, Authorship, citations, acknowledgments and visibility in social media: Symbolic capital in the multifaceted reward system of science. *Soc. Sci. Inf.* **57**, 223–248 (2018).
4. J. Ennser-Kananen, Are we who we cite? On epistemological injustices, citing practices, and #metoo in academia. *Apples - J. Appl. Lang. Stud.* **13**, 65–69 (2019).
5. A. F. Selvi, The myopic focus on decoloniality in applied linguistics and English language education: citations and stolen subjectivities. *Appl. Linguist. Rev.* **16**, 137–161 (2024).
6. P. M. Davis, Reward or persuasion? The battle to define the meaning of a citation. *Learn. Publ.* **22**, 5–11 (2009).
7. F. Vasen, N. F. Sarthou, S. A. Romano, B. D. Gutiérrez, M. Pintos, Turning academics into researchers: The development of National Researcher Categorization Systems in Latin America. *Res. Eval.* **32**, 244–255 (2023).
8. D. Kozlowski, V. Larivière, C. R. Sugimoto, T. Monroe-White, Intersectional inequalities in science. *Proc. Natl. Acad. Sci.* **119**, e2113067119 (2022).
9. C. Pradier, D. Kozlowski, N. S. Shokida, V. Larivière, Science for whom? The influence of the regional academic circuit on gender inequalities in Latin America. *J. Assoc. Inf. Sci. Technol.* 1–13 (2024). https://doi.org/10.1002/asi.24972.
10. J. J. Tóth, G. Háló, M. Goyanes, Beyond views, productivity, and citations: measuring geopolitical differences of scientific impact in communication research. *Scientometrics* **128**, 5705–5729 (2023).
11. T. Amano, *et al.*, The manifold costs of being a non-native English speaker in science. *PLOS Biol.* **21**, e3002184 (2023).
12. P. Mongeon, A. Paul-Hus, The journal coverage of Web of Science and Scopus: a comparative analysis. *Scientometrics* **106**, 213–228 (2016).
13. M. Salatino, Linguistic circuits of Latin American scientific production. *Tempo Soc.* **34**, 253–294 (2022).
14. M.-A. Vera-Baceta, M. Thelwall, K. Kousha, Web of Science and Scopus language coverage. *Scientometrics* **121**, 1803–1813 (2019).
15. Helsinki Initiative on Multilingualism in Scholarly Communication. (2019).
16. F. Beigel, "Multilingüismo y bibliodiversidad en América Latina" in *Anuario de Glotopolítica Núm. 5*, D. Bentivegna, J. del Valle, M. Niro, L. Villa, Eds. (Tipográfica, 2022), pp. 119–132.
17. R. Cruz Romero, D. Stephen, S. Stahlschmidt, Assessing bibliodiversity through reference lists: A text analysis approach. (2024).
18. F. Navarro, *et al.*, Rethinking English as a lingua franca in scientific-academic contexts: A position statement. *J. Engl. Res. Publ. Purp.* **3**, 143–153 (2022).
19. J. Shi, Articulations of Language and Value(s) in Scholarly Publishing Circuits. *Can. J. Acad. Librariansh.* **9**, 1–33 (2023).
20. G. Sivertsen, Multilingüisme equilibrat en ciència. *BiD Textos Univ. Bibl. Doc.* (2018). https://dx.doi.org/10.1344/BiD2018.40.24.
21. K. Gong, J. Xie, Y. Cheng, V. Larivière, C. R. Sugimoto, The citation advantage of foreign language references for Chinese social science papers. *Scientometrics* **120**, 1439–1460 (2019).
22. S. Hanafi, R. Arvanitis, The marginalization of the Arab language in social science: Structural constraints and dependency by choice. *Curr. Sociol.* **62**, 723–742 (2014).
23. E.-Y. J. Kim, Scholarly Publishing in Korea: Language, Perception, Practice of Korean University Faculty. *Publ. Res. Q.* **34**, 554–567 (2018).
24. N. Smirnova, T. Lillis, Citation in global academic knowledge making: A paired text history methodology for studying citation practices in English and Russian. *J. Engl. Res. Publ. Purp.* **3**, 78–108 (2022).
25. R. Waast, R. Arvanitis, C. Richard-Waast, P. L. Rossi, "What do social sciences in North African countries focus on ?" in *World Social Science Report*, (UNESCO, 2010), pp. 176–180.
26. E. Garfield, A. Welljams-Dorof, Language Use in International Research: A Citation Analysis. *Ann. Am. Acad. Pol. Soc. Sci.* **511**, 10–24 (1990).

27. T. Asubiaro, S. Onaolapo, D. Mills, Regional disparities in Web of Science and Scopus journal coverage. *Scientometrics* **129**, 1469–1491 (2024).
28. M. Huskisson, Guest Post: Reflections from The Munin Conference Part Three – Measuring Impact. *Sch. Kitchen* (2025). Available at: https://scholarlykitchen.sspnet.org/2025/01/23/guest-post-reflections-from-the-munin-conference-part-three-measuring-impact/ [Accessed 25 February 2025].
29. A. Bookstein, M. Yitzhaki, Own-language preference: A new measure of "relative language self-citation." *Scientometrics* **46**, 337–348 (1999).
30. F. Beigel, Les revues argentines de sciences humaines et sociales, entre la circulation régionale et l'ancrage local. *Biens Symb. Symb. Goods Rev. Sci. Soc. Sur Arts Cult. Idées* (2023). https://doi.org/10.4000/bssg.1790.
31. M. Salatino, Circuitos locales en contextos globales de circulación. Una aproximación a las revistas científicas argentinas. *Palabra Clave Plata* **9**, e073–e073 (2019).
32. P. Bourdieu, *Science de la science et réflexivité: cours du Collège de France, 2000-2001* (Raisons d'agir, 2001).
33. A. L. Packer, "The Pasts, Presents, and Futures of SciELO" in *Reassembling Scholarly Communications: Histories, Infrastructures, and Global Politics of Open Access*, M. P. Eve, J. Gray, Eds. (MIT Press, 2020), pp. 297–313.
34. M. Huskisson, Guest Post – Scholarly Publishing as a Global Endeavor: Leveraging Open Source Software for Bibliodiversity. *Sch. Kitchen* (2023). Available at: https://scholarlykitchen.sspnet.org/2023/02/16/guest-post-scholarly-publishing-as-a-global-endeavor-leveraging-open-source-software-for-bibliodiversity/ [Accessed 25 February 2025].
35. J. Heilbron, Y. Gingras, "The Globalization of European Research in the Social Sciences and Humanities (1980–2014): A Bibliometric Study" in *The Social and Human Sciences in Global Power Relations*, J. Heilbron, G. Sorá, T. Boncourt, Eds. (Springer International Publishing, 2018), pp. 29–58.
36. F. Beigel, M. Salatino, Circuitos segmentados de consagración académica: las revistas de Ciencias Sociales y Humanas en la Argentina. *Inf. Cult. Soc.* 11–36 (2015).
37. M. Salatino, Más Allá de la Indexación: Circuitos de Publicación de Ciencias Sociales en Argentina y Brasil. *Dados* **61**, 255–287 (2018).
38. J. P. Alperin, "The public impact of Latin America's approach to open access," Stanford University. (2015).
39. D. Chavarro, P. Tang, I. Ràfols, Why researchers publish in non-mainstream journals: Training, knowledge bridging, and gap filling. *Res. Policy* **46**, 1666–1680 (2017).
40. B. B. Kachru, World Englishes: approaches, issues and resources. *Lang. Teach.* **25**, 1–14 (1992).
41. I. Melliti, Une sociologie tunisienne francophone fait-elle encore sens ? *SociologieS* (2019). https://doi.org/10.4000/sociologies.9713.
42. S. Canagarajah, Language diversity in academic writing: toward decolonizing scholarly publishing. *J. Multicult. Discourses* **17**, 107–128 (2022).
43. S. Canagarajah, Diversifying "English" at the Decolonial Turn. *TESOL Q.* **59**, 378–389 (2025).
44. S. Khanna, J. Ball, J. P. Alperin, J. Willinsky, Recalibrating the scope of scholarly publishing: A modest step in a vast decolonization process. *Quant. Sci. Stud.* **3**, 912–930 (2022).
45. L. Céspedes, *et al.*, Evaluating the linguistic coverage of OpenAlex: An assessment of metadata accuracy and completeness. *J. Assoc. Inf. Sci. Technol.* 1–12 (2025). https://doi.org/10.1002/asi.24979.
46. U. Ammon, Language Planning for International Scientific Communication: An Overview of Questions and Potential Solutions. *Curr. Issues Lang. Plan.* **7**, 1–30 (2006).
47. M. D. Gordin, *Scientific Babel: How Science Was Done Before and After Global English* (University of Chicago Press, 2015).

48. M. Kuteeva, Knowledge flows and languages of publication: English as a bridge and a fence in international knowledge exchanges. *J. Engl. Res. Publ. Purp.* **4**, 80–93 (2023).
49. F. Beigel, Cartographies for an inclusive Open Science. [Preprint] (2024). Available at: https://preprints.scielo.org/index.php/scielo/preprint/view/10286 [Accessed 30 October 2024].
50. C. N. Authier, Cobertura de los identificadores persistentes (PID) para artículos científicos en Latinoamérica. [Preprint] (2023).
51. J. Law, W. Lin, Provincializing Sts: Postcoloniality, Symmetry, and Method. *East Asian Sci. Technol. Soc. Int. J.* **11**, 211–227 (2017).
52. M. J. Curry, T. M. Lillis, Strategies and tactics in academic knowledge production by multilingual scholars. *Educ. Policy Anal. Arch.* **22**, 32–32 (2014).
53. Y. Gingras, M. Khelfaoui, The relative (in)visibility of sociologists in the French, American, British, and German national fields (1970–2018). *Soc. Sci. Inf.* 05390184251317451 (2025). https://doi.org/10.1177/05390184251317451.
54. L. Céspedes, Latin American journals and hegemonic languages for academic publishing in Scopus and Web of Science. *Trab. Em Linguística Apl.* **60**, 141–154 (2021).
55. P. Bourdieu, Les conditions sociales de la circulation internationale des idées. *Actes Rech. En Sci. Soc.* **145**, 3–8 (2002).
56. K. Bennett, "Towards an epistemological monoculture: Mechanisms of epistemicide in European research publication" in *English as a Scientific and Research Language: Debates and Discourses*, R. Plo Alastrué, C. Pérez-Llantada, Eds. (De Gruyter Mouton, 2015), pp. 9–35.
57. C. Herzog, D. Hook, S. Konkiel, Dimensions: Bringing down barriers between scientometricians and data. *Quant. Sci. Stud.* **1**, 387–395 (2020).
58. J. Priem, H. Piwowar, R. Orr, OpenAlex: A fully-open index of scholarly works, authors, venues, institutions, and concepts. [Preprint] (2022). Available at: http://arxiv.org/abs/2205.01833 [Accessed 18 March 2025].

**Supporting Information for**

*A smack of all neighbouring languages*: How multilingual is scholarly communication

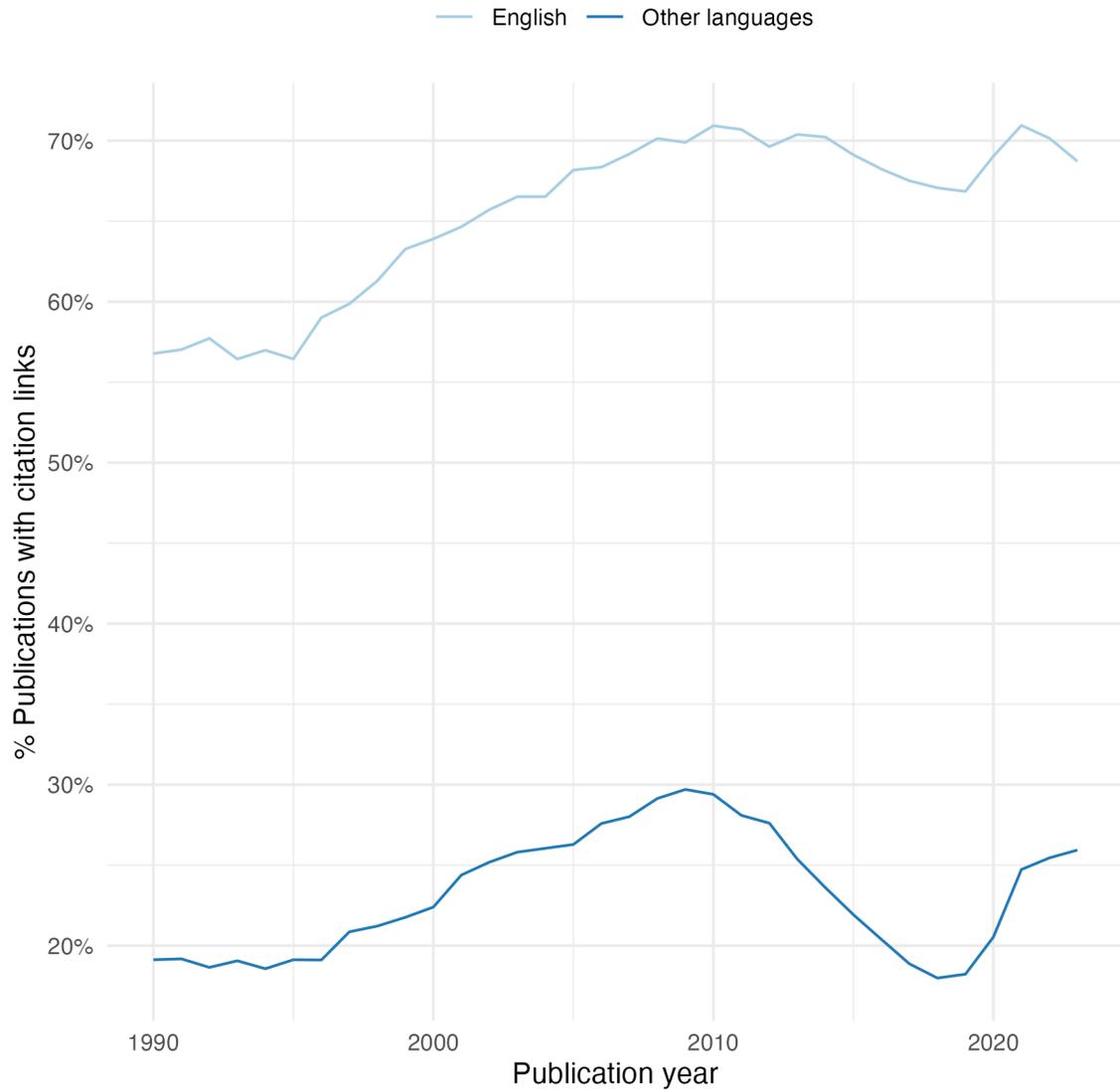

**Fig. S1.** Percentage of publications in Dimensions (1990-2023) with and without citation links according to language of publication (English / non-English).

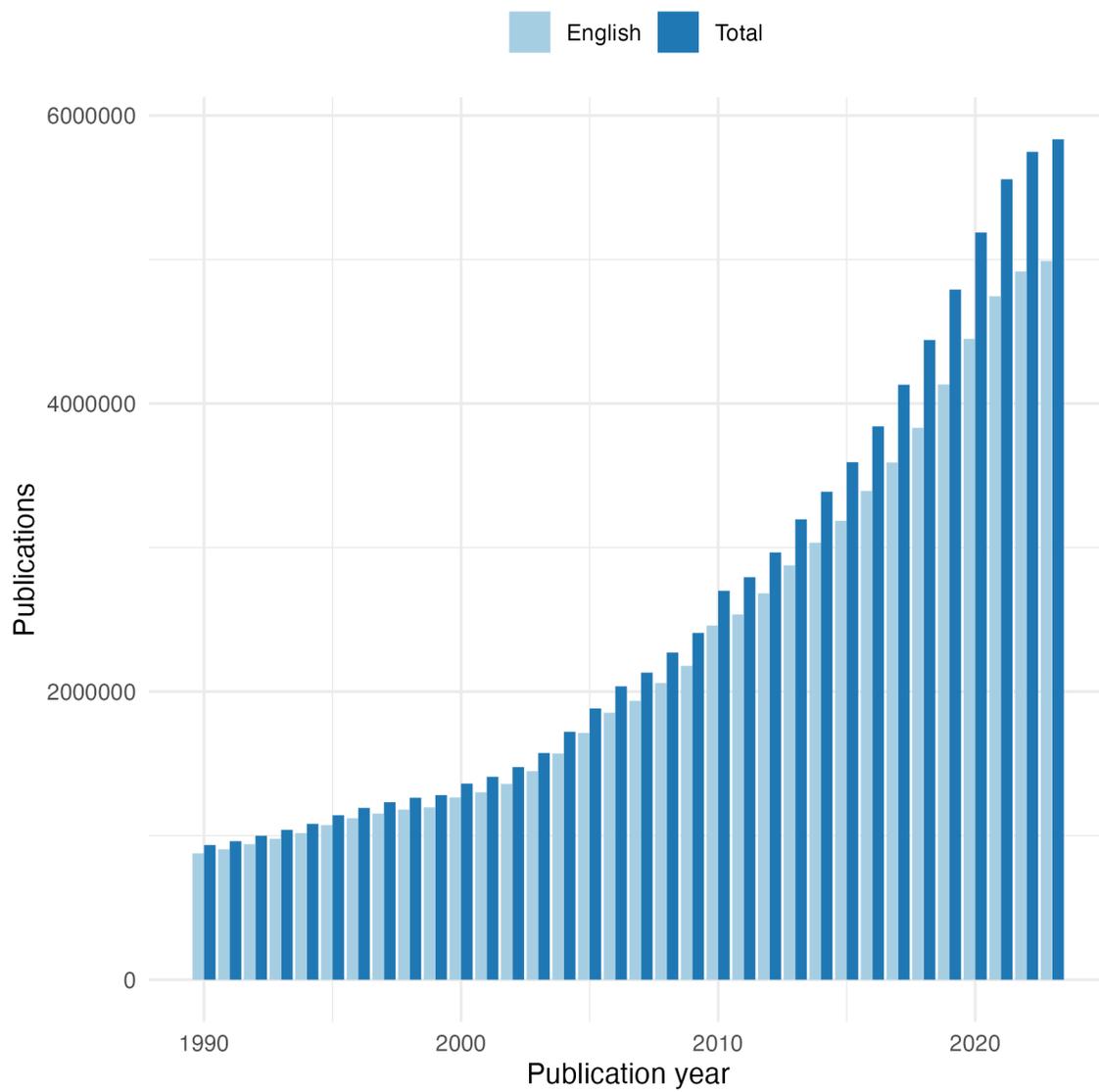

**Fig. S2.** Number of published articles and conference proceedings in Dimensions (1990-2023), in English and in total.

**Software S1. Interactive and extended results.** The interactive and extended version of our results is publicly available under: https://vlab.ebsi.umontreal.ca/languages_app/

**Software S2. Scripts.** The scripts used for this study are publicly available under: https://github.com/caropradier/languages_science